# $Cu_2O$ as nonmagnetic semiconductor for spin transport in crystalline oxide electronics

I.Pallecchi [a], L.Pellegrino [a], N. Banerjee [c], M. Cantoni [d], A.Gadaleta [a,b], A.S.Siri [a,b], D.Marré [a,b]

[a] *CNR-SPIN, Corso Perrone 24, 16152 Genova, Italy*
[b] *Università di Genova, Dipartimento di Fisica, Via Dodecaneso 33, 16146 Genova, Italy*
[c] *Department of Materials Science, Cambridge University, Pembroke Street, Cambridge CB2 3QZ, United Kingdom*
[d] *L-NESS, Dipartimento di Fisica - Politecnico di Milano, Via Anzani 42, 22100 Como, Italy*

## Abstract

We probe spin transport in $Cu_2O$ by measuring spin valve effect in $La_{0.7}Sr_{0.3}MnO_3/Cu_2O/Co$ and $La_{0.7}Sr_{0.3}MnO_3/Cu_2O/La_{0.7}Sr_{0.3}MnO_3$ epitaxial heterostructures. In $La_{0.7}Sr_{0.3}MnO_3/Cu_2O/Co$ systems we find that a fraction of out-of-equilibrium spin polarized carrier actually travel across the $Cu_2O$ layer up to distances of almost 100 nm at low temperature. The corresponding spin diffusion length $d_{spin}$ is estimated around 40 nm. Furthermore, we find that the insertion of a $SrTiO_3$ tunneling barrier does not improve spin injection, likely due to the matching of resistances at the interfaces. Our result on $d_{spin}$ may be likely improved, both in terms of $Cu_2O$ crystalline quality and sub-micrometric morphology and in terms of device geometry, indicating that $Cu_2O$ is a potential material for efficient spin transport in devices based on crystalline oxides.

## 1. Introduction

The main goal of spintronics is predicting, fabricating and analyzing various device architectures that can be controlled by electric and magnetic fields, thus adding new functionalities to conventional electronics devices. The spin is conserved in most of the scattering events and decays within a characteristic time $\tau_{spin} >> \tau$ ($\tau$ mean scattering time). Indeed, only the spin dependent part of the scattering potential is responsible for the spin decay (e.g. the spin-orbit potential). For example $\tau/\tau_{spin} \sim 10^{-3}$ in Cu and Al at low temperature [1]. This indicates that spintronics could allow to carry information for much longer times as compared to electronics. A large number of the proposed devices are based on a semiconducting nonmagnetic element where non equilibrium spin polarized carriers are injected [2,3,4,5]. Among these, the Datta and Das spin transistor [6] is still considered the paradigm among spintronics devices. Not all the proposed devices have been actually fabricated with satisfactory performances. The limiting obstacles yet to be overcome are mainly related to the most suitable materials employed, both as ferromagnetic and nonmagnetic elements.

Spin injection have been tried in superconductors [7], metals [1,8], semiconductors [9], as well as in organic semiconductors[10] and carbon nanotubes [11]. In particular, spin injection in semiconductors is of particular interest, due to the versatility of semiconductors in terms of doping, micro- and nano-structures fabrication, tuning of optical properties and spin-orbit coupling, bipolar (electrons and holes) transport and above all due to the possibility of integration with conventional electronics. Moreover, in semiconductors the scattering times are longer (despite Fermi velocities are smaller); thereby, longer $\tau_{spin}$ are expected, even if not necessarily larger spin coherence lengths $d_{spin}$. Spin coherence for distances beyond 100 μm have been observed for optically excited spin in a semiconductor [12] and in any case, apart from such record values, typically observed values are $d_{spin} \sim 1 \mu m$ for Cu, Ag and Al [13] and $d_{spin} \sim$ several μm for conventional semiconductors [14].

Non-equilibrium spin injection can be realized by optical methods or via transport. In the latter case, injection from ferromagnetic electrodes presents several drawbacks. With ferromagnetic transition metal electrodes such as Co and Fe, the problem of resistance mismatch, which is responsible for severe depolarization at the interface, has to be faced [15,16],. In other words, as a consequence of the different conductivity of the two materials at either sides of the interface, charge continuity equation and charge quasi-neutrality conditions yield a voltage build-up at the interface, proportional to the spin accumulation therein. The meaning of this interface resistance is that the spin accumulation and the finite spin relaxation time act as a bottleneck for carrier injection, as a consequence of the fact that the spin carriers are also charge carriers. The problem of resistance mismatch could be circumvented by choosing a more resistive ferromagnetic electrode. On the other hand, more resistive semiconductors doped with magnetic ions such as (In,Mn)As, (Ga,Mn)As, (Zn,Cr)Te present some drawbacks: they are only hole doped, their Curie temperature is usually much below room temperature (~200K) and their spin polarization is too low for them to be considered as potential ferromagnetic injecting electrodes.

Actually, for spin injection via transport from ferromagnetic electrodes, silicon would be a much more appealing nonmagnetic semiconductor, as it could be the “all-in-one” material, due to its low cost and predominant role in conventional electronics. However, a satisfactory spin injection in silicon has not been achieved until very recently [17,18]. Clearly the search of the fittest nonmagnetic semiconductor is still an open challenge. The target is an high mobility semiconductor, with fairly low spin-orbit coupling and possibility of epitaxial growth with ferromagnetic metallic and nonmagnetic insulating layers, so as to realize high quality interfaces.

In this context, an unconventional route to explore is the world of isostructural oxides, where possible candidates for the roles of ferromagnetic injecting electrodes and nonmagnetic semiconductors have to be found. Recently, the possibility of building an unconventional electronics fully based on epitaxial oxides had been put forward. On one hand, intrinsic limits of silicon-based electronics could be overcome, allowing a higher degree of integration density, and on the other hand novel devices made of functional materials could be fabricated. Three-dimensional integration of perovskites could allow to combine high-$T_c$ superconductivity of cuprates, half-metallic almost 100% spin polarized transport and colossal magnetoresistance of manganites, ferroelectricity and piezoelectricity of titanates, multiferroic coupling of $YMnO_3$ and $BiMnO_3$.

Manganites as injecting electrodes are a natural choice, at least at low temperatures, thanks to their spin polarization of almost 100% and Curie temperatures around room temperature. Despite the issue of depolarization at the interface is a usual drawback for manganites, it has been tackled and tunneling magnetoresistance values larger than 1800% at 4K, corresponding to polarization ~95%, have been measured in spin valve heterostructures [19]. However, *no attempt has yet been made of injecting out-of-equilibrium spin polarized carriers into a nonmagnetic crystalline oxide semiconductor* and studying spin polarized transport therein, inside an oxide epitaxial heterostructure. Such kind of device could open the way to a whole class of crystalline oxide spintronics devices, which could be also integrated with oxide electronics devices and represent a valuable alternative to spintronics devices based on standard semiconductors.

Finding an all-purpose nonmagnetic semiconducting oxide to build unconventional electronics and spintronics is the first and most difficult challenge. It should have high mobility and bipolar transport and it should be structurally and chemically compatible with other perovskite oxides, especially manganites.

Recently, the possibility of growing epitaxial and oriented $Cu_2O$ films on perovskite $SrTiO_3$ substrates has been evidenced [20]. Cuprous oxide $Cu_2O$ is a p-type semiconductor, with a direct optical bandgap of 1.9-2.1 eV [21,22] and an effective mass around $0.84 \cdot m_0$ [22]. Its low average atomic number points to a low spin-orbit coupling. As for its crystal structure, the oxygen atoms form a bcc lattice with cubic lattice parameter $a \approx 4.27Å$, while the copper atoms are on the vertices of a tetrahedron around each oxygen atom. $Cu_2O$ has already been employed in the fabrication of electronic devices, thanks to its low cost, non-toxicity, fairly good carrier mobility, high minority carriers diffusion length, direct energy gap; for example it has been used in film solar cells[23], photovoltaic devices[24], resistive switching memories[25] and transistors[26].

It appears that spin diffusion from manganite electrodes into $Cu_2O$ could be a remarkable bet, also considering that a favorable band alignment and a low resistance mismatch should not cause appreciable spin depolarization at the interface. Moreover, $Cu_2O$ may be used also as p-type semiconducting element within oxide electronics, which is a still vacant role, while there are plenty of n-type semiconducting oxides available. In this work, we explore this possibility by measuring spin injection in $La_{0.7}Sr_{0.3}MnO_3$(LSMO)/$Cu_2O$/Co and LSMO/$Cu_2O$/ LSMO trilayers and extracting an estimation of the spin diffusion length in $Cu_2O$.

## 2. Experimental

We deposit LSMO/$Cu_2O$/Co and LSMO/$Cu_2O$/LSMO trilayers by pulsed laser ablation on $SrTiO_3$ (001) substrates. In situ shadow masking is used to allow electrical contacts to the bottom layers. The compatibility of growth conditions of the different elements in a key issue. Deposition of manganites requires high temperatures and high oxygen pressures, whereas the $Cu_2O$ layer requires moderate temperatures and low oxygen pressures, in order to avoid formation of CuO secondary phase. This makes the deposition of the uppermost LSMO layer critical, as the underlying $Cu_2O$ layer just deposited must not be turned into CuO.

LSMO is deposited at 815°C substrate temperature, $5 \cdot 10^{-2}$ Torr oxygen pressure, 2Hz laser repetition rate and 1.2 J/cm$^2$ laser energy density, corresponding to a deposition rate of 0.07Å/pulse. The bottom LSMO layer is post-annealed for half an hour at 600°C and 200 Torr. $Cu_2O$ is deposited at 650°C substrate temperature, $5 \cdot 10^{-4}$ Torr oxygen pressure, 5Hz laser repetition rate and 1.2 J/cm$^2$ laser energy density, corresponding to a deposition rate of 0.027Å/pulse. In the case of the upper LSMO layer, the oxygen pressure is raised to $5 \cdot 10^{-2}$ Torr at the very last moment and no post annealing is carried out, while all the other parameters are the same as for the bottom LSMO electrode. Co is deposited at room temperature and high vacuum ($5 \cdot 10^{-7}$ Torr background pressure),

10Hz laser repetition rate and 1.2 $J/cm^2$ laser energy density, corresponding to a deposition rate of 0.083Å/pulse.

In the case of LSMO/$Cu_2O$/Co trilayers, we also explore the effect of a tunneling $SrTiO_3$ barrier for spin injection, deposited between the ferromagnetic electrodes and the $Cu_2O$ layer, thus obtaining LSMO/$SrTiO_3$/$Cu_2O$/$SrTiO_3$/Co and LSMO/$SrTiO_3$/$Cu_2O$/Co heterostructures. In both cases, the $SrTiO_3$ layer sandwiched between LSMO and $Cu_2O$ is deposited in the same conditions as the manganites, thus turning out perfectly epitaxial and oriented. On the contrary, the upper $SrTiO_3$ layer sandwiched between $Cu_2O$ and Co is deposited at room temperature and $5 \cdot 10^{-4}$ Torr oxygen pressure to avoid oxidation of the underlying $Cu_2O$ layer, thus turning out amorphous and with many defects and traps. Finally, as a reference to check the behavior in presence of electrical shorts, LSMO/Co heterostructures are prepared. Single layer $Cu_2O$ films are also deposited for in-plane transport characterization.

The heterostructures are characterized by X-rays diffraction, in order to study phase formation and purity, structural quality and epitaxy. The surface morphology of different materials are inspected by atomic force microscopy (AFM). Magnetization measurements as a function of applied field and temperature are carried out in a SQUID magnetometer by Quantum Design up to 5T. Transport properties and spin injection are measured in a Quantum Design Physical Property Measurement System (PPMS), from 10K to room temperature and in magnetic fields up to 9T. Current-voltage characteristics, resistance versus cycled magnetic field, Hall effect and resistance versus temperature measurements are carried out on single films and heterostructures. In the case of vertical resistance versus cycled magnetic field of trilayers, the magnetic field is applied parallel to the interfaces, that is in the plane where the easy magnetization axes of the electrodes lie. In the case of Hall effect measurements on $Cu_2O$, the magnetic field is applied perpendicularly to the film plane and the film is patterned in the shape of a Hall bar by optical lithography and wet etching in HCl.

## 3. Results and discussion

### 3.a. Structural and morphological characterization

In figure 1, X-rays patterns of a LSMO/$Cu_2O$/LSMO trilayer is shown. It can be seen that single phase and c-oriented LSMO and $Cu_2O$ are present. In fact, in a previous work we have demonstrated cube-on-cube growth of $Cu_2O$ on perovskite oxides [20], despite the significant mismatch (9.5%) between the lattice constants of the bulk $Cu_2O$ (*a≈4.27Å)* and of the $SrTiO_3$ substrate (*a≈3.905Å)*. Similar patterns are obtained also for LSMO/$Cu_2O$/Co trilayers. The out-of-plane lattice parameter of the $Cu_2O$ layer turns out to be 4.31Å, slightly larger than the bulk value.

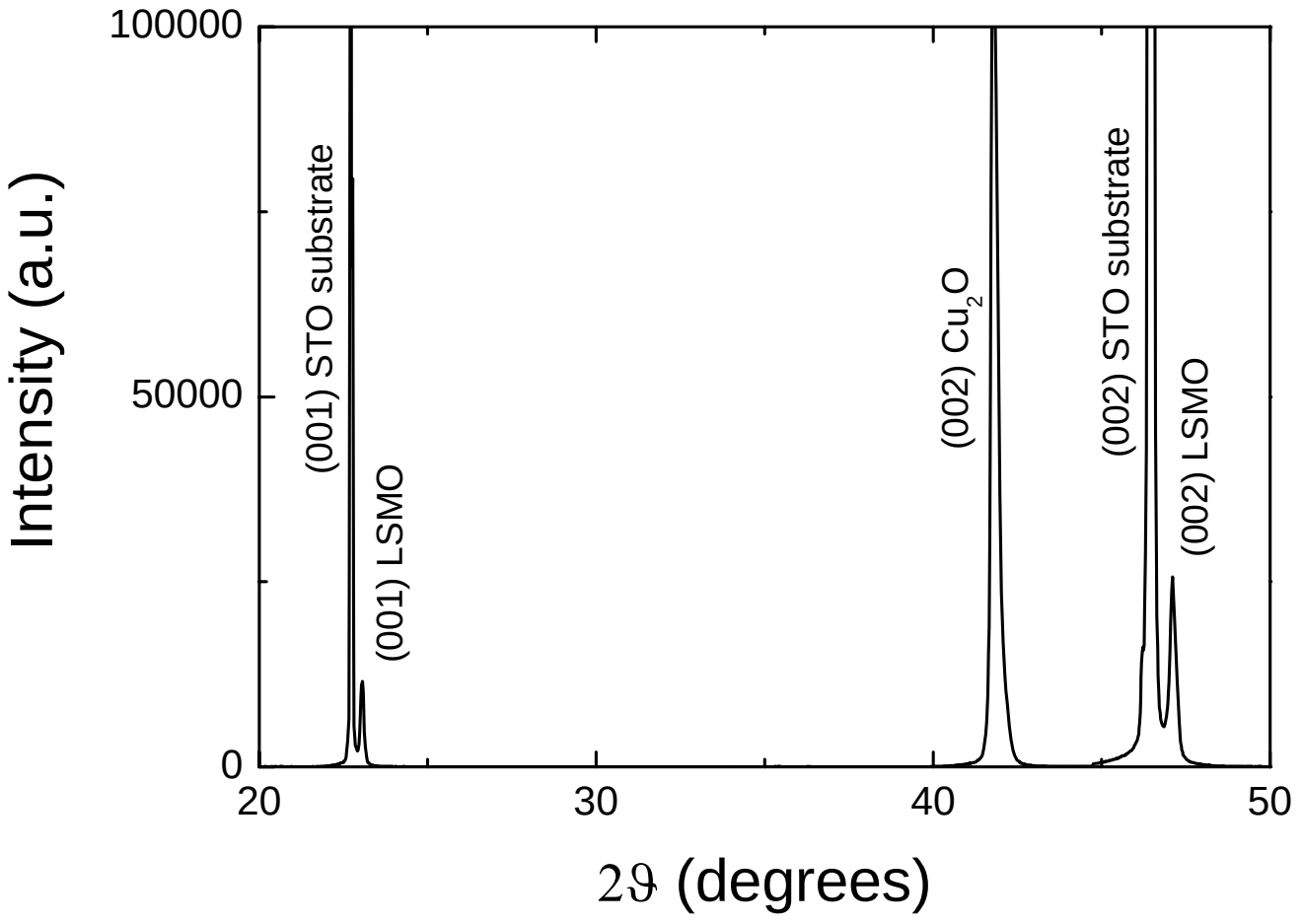

**Figure 1:** X-rays pattern of a LSMO/$Cu_2O$/LSMO trilayer with 1.2μm thick $Cu_2O$.

As surface smoothness at atomic level is crucial for high quality interfaces and thus for preservation of spin polarization, AFM analysis of LSMO film surfaces is carried out. In figure 2a), the image of a 200 nm thick film shows well visible atomic terraces and a root mean square roughness as low as 0.36 nm r.m.s., despite the considerable thickness. Particulates whose height is at most 6nm can be seen. Absence of sizeable particulate and atomic flatness are necessary prerequisites for fabrication of pinhole-free heterostructures. On the contrary, the morphology of a LSMO film deposited on the top of a several hundredths nm thick $Cu_2O$ film exhibits a granular structure and a surface roughness of 20 nm r.m.s., as shown in figure 2b). While hundredths nm thick $Cu_2O$ films, although epitaxial and c-oriented, present the same rough morphology of figure 2b), with decreasing thickness much smoother and voidless samples are obtained. In figure 2c), the morphology of a 25 nm thick $Cu_2O$ film is shown: this film has a much smaller surface roughness of few nm, even if grains of average size ~100nm can be found.

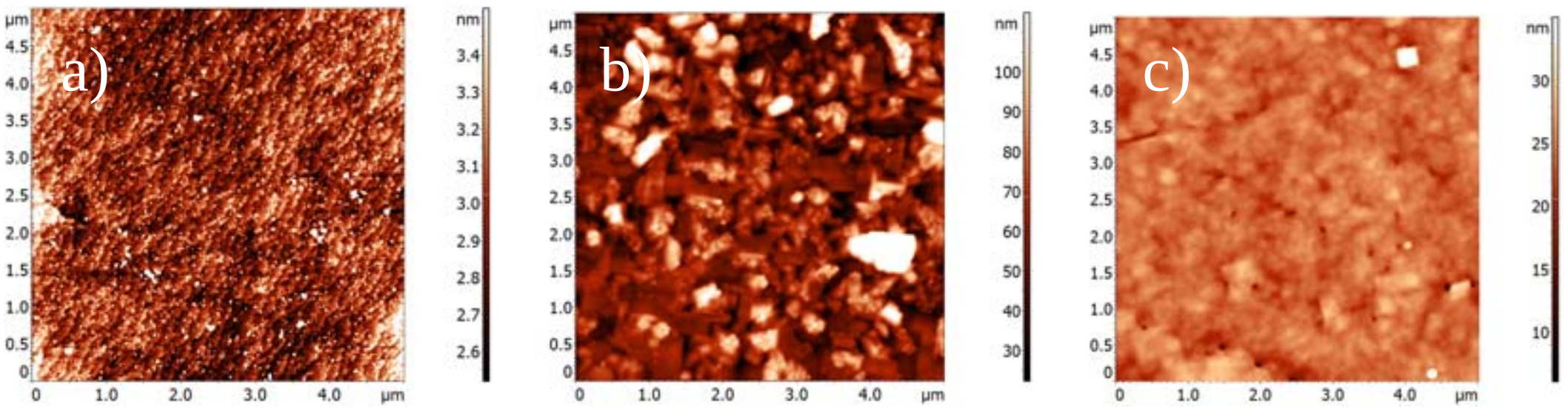

**Figure 2** (color online)**:** a) 5μm X 5μm AFM image of a 200 nm thick LSMO film; the surface is atomically smooth and atomic terraces are clearly seen. b) 5μm X 5μm AFM image of a LSMO film deposited on the top of a $Cu_2O$ several hundredths nm thick film, exhibiting a granular structure and a surface roughness of 20 nm r.m.s. c) 5μm X 5μm AFM image of a 25nm thick $Cu_2O$ film, exhibiting a surface roughness of 3 nm r.m.s.

## 3.b. Electrical transport across $Cu_2O$

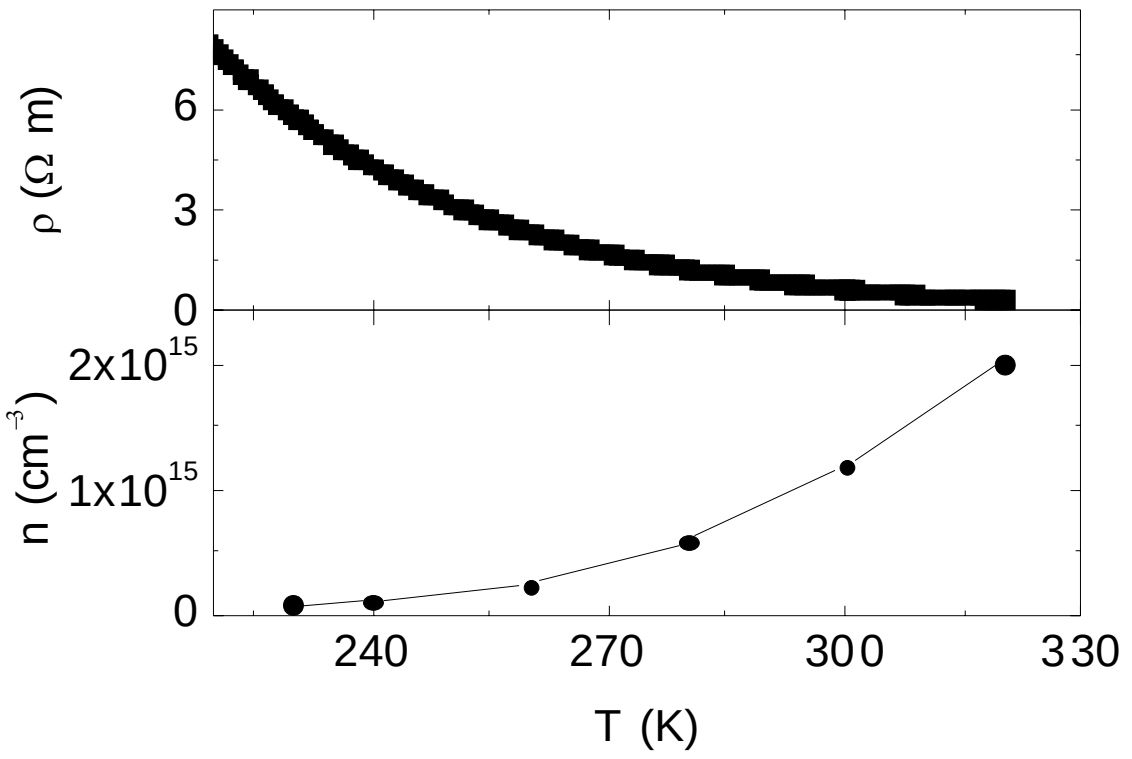

**Figure 3:** Resistivity and carrier concentration measured by Hall effect in a 150 nm thick $Cu_2O$ film, patterned as a 50 μm wide Hall bar.

In figure 3, the electrical characterization of a 150 nm thick $Cu_2O$ film patterned as an Hall bar is shown. The resistivity $\rho$ increases exponentially with decreasing temperature, with a room temperature value $\rho(300K)\approx 0.72\ \Omega m$. The carrier concentration $n$ also exhibits an exponential thermally activated behavior, with a room temperature value $n(300K)\approx 1.2\cdot 10^{15}\ cm^{-3}$. Thereby, the Hall mobility is fairly good, $\mu_H \approx 70\ cm^2/(Vs)$ at room temperature, in agreement with literature values [23]. The activation energy turns out to be $\Delta\approx 0.2eV$. Extrapolations to 10K yield $\rho(10K)\approx 4\cdot 10^3\ \Omega m$ and $n(10K)\approx 2.8\cdot 10^{10}\ cm^{-3}$. Using a free electron picture, the electron mean free $l$ path can be extracted as $l = \left(3\pi^2 n\right)^{1/3} \frac{\hbar}{n\rho e^2}$, where $\hbar = h/2\pi$, $h$ is the Planck constant and $e$ is the electron charge. We obtain values of the order of the lattice spacing, weakly dependent on temperature.

### 3.c. Probe of spin transport across $Cu_2O$

Turning now to spin injection experiments, the band alignment between LSMO and $Cu_2O$ must be considered. The electron affinity $\chi$ measured in $Cu_2O$ films is 2.9 eV (3.2 eV measured in bulk samples) [27], the bandgap $E_{gap}$ is 1.9 eV – 2.0 eV [21,22] and typically the Fermi level lies $h$=0.45 eV above the top of the valence band [28] (a value $h$=0.25 eV has been also reported [28, 29], in better agreement with the activation energy extracted from resistivity and carrier concentration data of figure 3); hence, it turns out that the workfunction $\Phi=\chi+E_{gap}-h$ is in the range 4.3 eV – 4.7 eV. If this value is compared for example with the slightly larger value of perovskite manganites, which is 4.7 eV- 4.9 eV[30,31,32], there may be the conditions for a upward bending of $Cu_2O$ bands at the $Cu_2O$/LSMO interface, and thereby for diffusion of spin polarized holes between manganites and $Cu_2O$ at finite temperatures. Indeed, the barrier height for such diffusion turns out to be ~0.2eV, similar to the activation energy of carriers in $Cu_2O$ extracted from the data of figure 3. This situation is depicted in figure 4. On the other hand, given the uncertainty on these estimations, especially the one on the position of the $Cu_2O$ Fermi level, the barrier may be much higher than 0.1-0.2eV, and in this case an insulating tunneling barrier between $Cu_2O$ and LSMO layers could be beneficial for spin injection by application of a voltage, rather than by simple diffusion. A direct measurement of band alignment by XPS (X-rays photoemission spectroscopy) technique is underway [33]. A similar argument is valid for the $Cu_2O$/Co interface, as the Co workfunction is around 5eV [34], not much different from that of the LSMO.

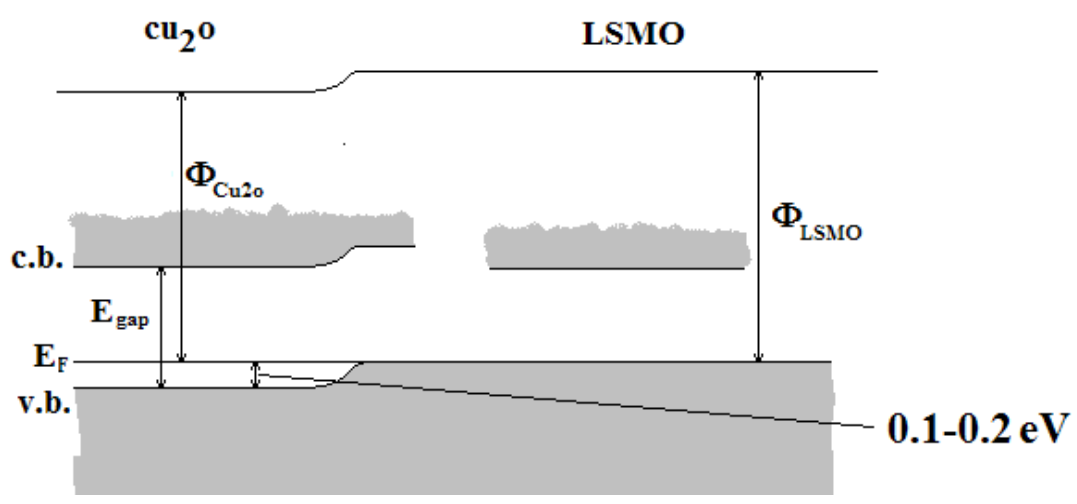

**Figure 4:** Tentative sketch of band alignment at the $Cu_2O$/LSMO interface. Φ, $E_{gap}$, $E_f$, c.b., v.b. indicate the workfunction, band gap, Fermi level, conduction band and valence band, respectively. The energy barrier for diffusion of holes from LSMO to $Cu_2O$, supposed to be around 0.1-0.2 eV, is also indicated.

Let now consider the resistance versus field behavior of a ferromagnet/$Cu_2O$/ferromagnet vertical geometry, like the one sketched in the inset of figure 5, with the field parallel to the layers. Provided that the spin polarization is maintained across the $Cu_2O$ layer, at least partially, a spin valve behavior is expected: the polarized carriers injected from one electrode reach the other electrode and depending on the relative orientations of electrode magnetizations, higher or smaller resistance values are measured. If such experiment is carried out on trilayers of different $Cu_2O$ thicknesses, an estimation of the spin diffusion length in $Cu_2O$ can be obtained.

In figure 5, we present representative measurements on a LSMO/$Cu_2O$/LSMO trilayer, where the $Cu_2O$ thickness is 1.6 μm (a similar behavior is observed also in samples with smaller $Cu_2O$ thickness). In the two left panels, the vertical measurements at temperatures 10K and 300K are reported, respectively. Clearly, beside the reversible negative magnetoresistivity of magnetic origin, there is a well visible spin valve hysteresis, much more pronounced at 10K but still visible at 300K. The hysteretic curves do not show abrupt jumps corresponding to magnetization switching; instead, the increasing-|H| and decreasing-|H| branches merge smoothly, and the increasing-|H| branches present two almost symmetric maxima at characteristic fields $\pm H_c$. These maxima originate from the resistance increase at the lowest fields due to the anisotropic magnetoresistance (AMR) and the resistance decrease at higher fields due to the negative magnetoresistance related to the alignment of Mn spins. For the AMR in manganites, dependent on the angle between the current and the magnetization, we refer to the study carried out in ref. [35]. The curve shapes in the left panels of figure 5, with no abrupt resistance jumps, indicate that there are not two well defined coercive fields for the lower and upper LSMO electrodes, but rather that there is a distribution of coercive fields, so that the magnetization switches gradually with increasing field. In order to better explore this phenomenology, we also present similar resistance versus field measurements of the upper and lower LSMO electrodes, alone, at 10K, in the right panels of figure 5. Whereas the bottom electrode, grown onto the substrate, has a well defined and almost vanishing coercive field and thus exhibits no hysteresis, the top electrode, grown on the rough $Cu_2O$ surface, exhibits evident hysteresis. The granular structure observed in AFM images (see figure 2b)) is responsible for a distribution of coercive fields, which determines the resistance hysteresis. Each grain is a magnetic domain having its own coercive field and a weak magnetic coupling with adjacent domains, so that it is rotated by the external field quite independently from the adjacent domains. The tunneling current between adjacent non aligned domains determines the hysteresis. Unluckily, this hysteretic contribution adds in series to the vertical measurements, so that the hysteresis displayed in the left panels of figure 5 cannot be unambiguously attributed to charge carriers that cross the $Cu_2O$ layer maintaining their spin polarization. Indeed, no clear trend of increasing hysteresis with decreasing $Cu_2O$ thickness in a series of heterostructures of this kind is observed. Hence, in these systems, nothing can be concluded about the spin diffusion length in $Cu_2O$.

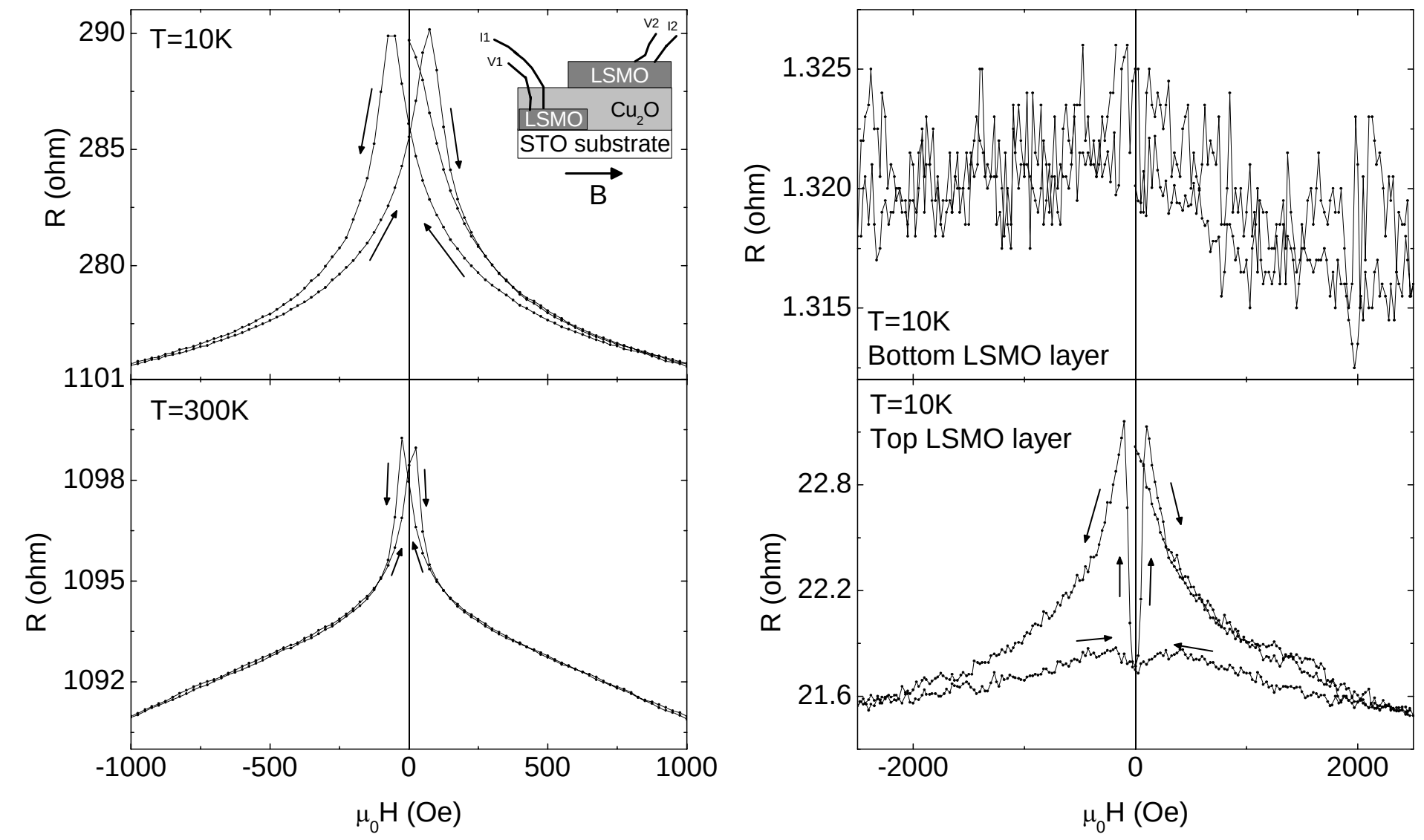

**Figure 5:** Left: resistance versus field curves of a LSMO/$Cu_2O$/LSMO trilayer with $Cu_2O$ thickness 1.6 μm, measured at 10K (upper panel) and 300K (lower panel). A sketch of the measurement configuration is shown in the inset. Right: resistance versus field curves of the bottom (upper panel) and top (lower panel) LSMO electrodes. The arrows indicate the direction of the magnetic field sweep.

On the contrary, LSMO/$Cu_2O$/Co trilayers may be helpful in this respect, as in the Co layer, even grown onto not atomically smooth surfaces, the adjacent magnetic domains are much more coupled, so that they remain almost parallel to each other when they are rotated by the external field. In other words, the distribution of domain orientations is much narrower and the resistance versus field curve measured on the upper Co electrode presents no hysteresis. The Co polarization, defined as the ratio of the density of states for majority and minority spin bands, is only 30-40% [36], as compared to the almost 100% polarization of manganites. By converse, the Co as ferromagnetic electrode has several advantages over LSMO, such as the less critical surface depolarization and the Curie temperature much larger than room temperature. Hence, despite the ultimate target of this work is the fabrication of an all-oxide planar or vertical device for spin injection, we now study the behavior of a LSMO/$Cu_2O$/Co trilayer in order to probe spin transport in $Cu_2O$.

In figure 6, we present resistance measurements on a LSMO/$Cu_2O$/Co trilayer, carried out in the same vertical configuration depicted in the inset of figure 5, with magnetic field parallel to the layers. The $Cu_2O$ thickness in this sample is 50nm. Current-voltage characteristics are ohmic at all temperatures and the temperature behavior is metallic, dominated by the contribution of the bottom LSMO electrode, as shown in the upper left panel of figure 6. This is not surprising, as, due to the geometrical factors, the $Cu_2O$ layer contributes with a series resistance that is 5 orders of magnitude smaller than the measured resistance, at room temperature. Its low temperature extrapolation is still a small fraction of the measured resistance, as well. Hence, the total resistance is dominated by the current path along the electrodes. This hinders a useful further check on the quantitative analysis of the hysteretic spin valve behavior, as discussed in the following.

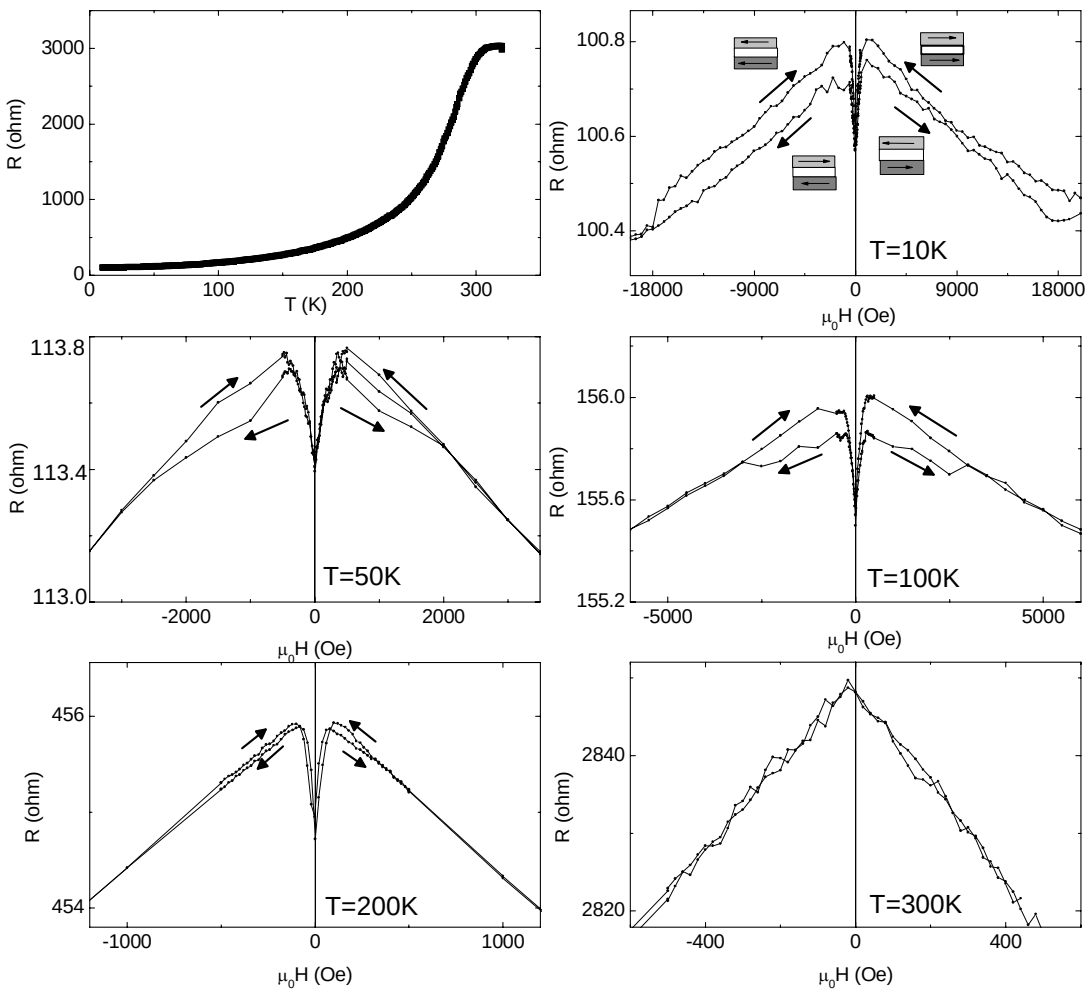

**Figure 6:** Upper left panel: resistance versus temperature across a LSMO/$Cu_2O$/Co trilayer with $Cu_2O$ thickness 50nm. Other panels: resistance versus magnetic field across the same trilayer, measured at different temperatures T=10K, 50K, 100K, 200K and 300K. The arrows indicate the direction of the magnetic field sweep. In the top left panel the directions of magnetizations of the lower LSMO and upper Co electrodes are also schematically indicated.

In the second to sixth panels of figure 6, resistance versus cyclically swept magnetic field on the same LSMO/$Cu_2O$/Co trilayer measured at different temperatures are shown. The shape of magnetoresistance curves has reversible contributions from the LSMO electrode, namely the negative term of magnetic origin, related to the alignment of spins by the external field, and the AMR term, related to the angle between the current and the magnetization. The dip at zero field is due to the AMR, as in this low field regime the resistance rises as the local magnetization is rotated perpendicular to the direction of the applied current by the external field. The resistance decrease at higher fields is due to the usual negative magnetoresistance of manganites. As a result of these terms, two symmetric resistance maxima are present in both increasing-|H| and decreasing-|H| branches. Disregarding these reversible contributions, we focus for our purposes on the hysteretic behavior, which is clearly seen in all curves below 100K, it is almost vanishing at 200K and it is completely disappeared in the 300K measurement. The two hysteretic lobes open up at the coercive field of the LSMO electrode, which is lower than 100Oe at 10K and even smaller at higher temperatures; indeed, the increasing-|H| and decreasing-|H| branches cross each other at this low field values. The field at which the increasing-|H| and decreasing-|H| branches of the curve merge is the same as that at which the Co electrode is fully aligned parallel to the applied field. Indeed, the upper Co electrode is certainly multidomain, as a consequence of its growth on the $Cu_2O$ surface, which is not atomically flat, so that the complete magnetic alignment is reached smoothly and only at large fields. This hypothesis is demonstrated by magnetization measurements M(H) carried out on LSMO/$Cu_2O$/Co heterostructures with in plane applied field, shown in figure 7. Actually, it can be seen that only at fields of the order of Teslas the magnetic hysteresis vanishes and the increasing-

|H| branches completely saturate, indicating that only at such high fields the Co electrode is fully aligned parallel to the applied field. This is consistent with the high field ~2T at which the resistance hysteresis loop closes (see figure 6). Noticeably, the same M(H) behaviour is observed at low and at room temperature, indicating that the Co electrode behaves similarly in this temperature range. Conversely, the resistance hysteresis is observed only below the Curie temperature of the LSMO layer (see figure 6), indicating that it comes from the simultaneous presence of both ferromagnetic Co and LSMO electrodes. This results, together with the fact that no hysteresis is observed if each single electrode is measured separately, demonstrates that the hysteresis is a *signature of the transfer of spin polarized carriers traveling across the $Cu_2O$ layer*. From the magnetic hysteresis loops of figure 7, it is also apparent that the coercive field of the Co layer is around 600 Oe at 10K (see inset). However, both magnetization and resistance hysteresis loops change smoothly rather than sharply as would occur in case of an abrupt monodomain switch. Hence, in order to track the magnetization direction of the electrodes, it is easier and more reliable to identify the point where the resistance hysteresis loop closes rather than the point corresponding to the coercive field, as usually occurs in spin valves.

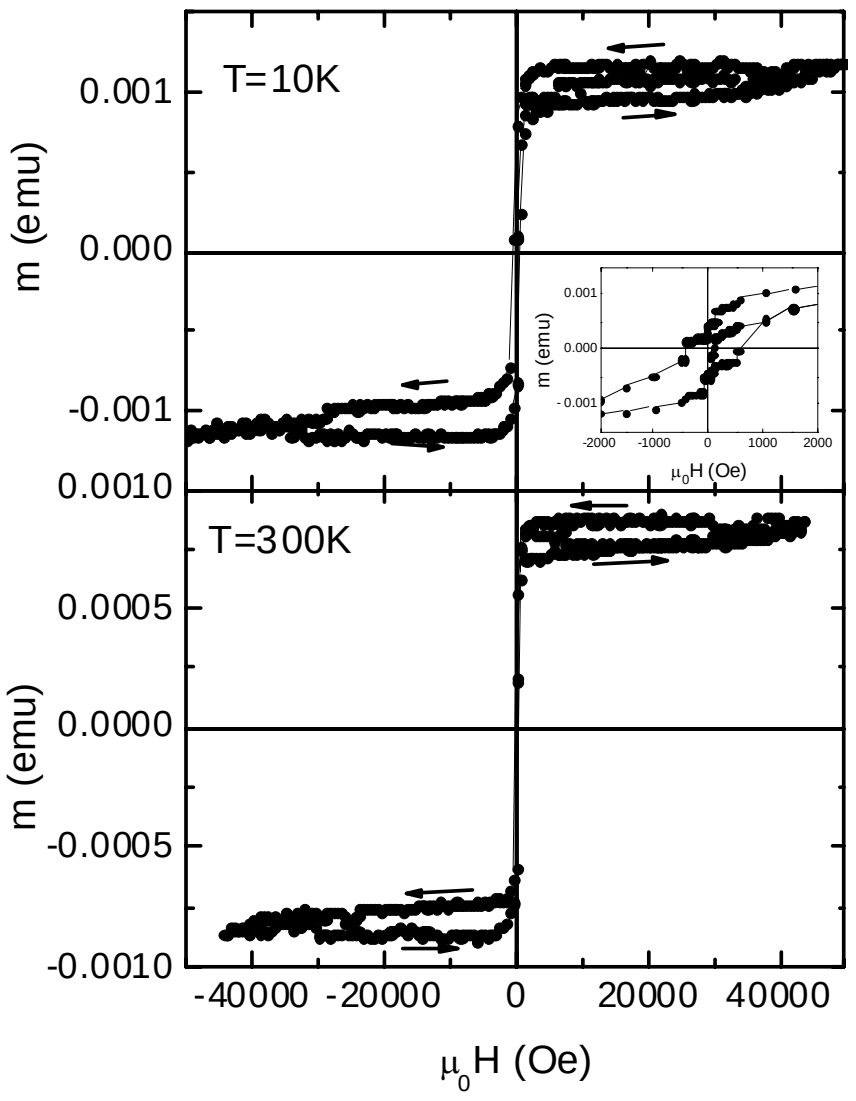

**Figure 7:** Magnetization measurements of a LSMO/$Cu_2O$/Co trilayer at 10K (upper panel) and 300K (lower panel). In the inset a zoom of the low field regime at 10 K is shown, showing a coercive field of the Co layer of around 600 Oe. The linear slope due to the diamagnetic contribution of the substrate has been subtracted. The arrows indicate the direction of the magnetic field sweep.

Turning back to figure 6, thanks to the large difference between the characteristic fields at which the increasing-|H| and decreasing-|H| branches of resistance curves depart and merge, especially at the lowest temperatures, the hysteresis is well visible. It can be seen that the resistance is larger when the magnetizations of the LSMO and Co electrodes are parallel (decreasing-|H| branches) and smaller when they are antiparallel (increasing-|H| branches). Just the opposite situation occurs when only LSMO electrodes are present (see figure 5). Indeed, as explained in ref. 36, in Co the density of states of the minority spin band is larger than that of the majority spin band, while in LSMO the opposite situation occurs. Hence, the transmission probability, expressed in terms of initial and final density of states, yields the observed magnetoresistance behavior. The same behavior is found in LSMO/$SrTiO_3$/Co magnetic tunnel junctions [36] and in structures with organic spacers between LSMO and Co electrodes [37]. We point out that in the measurements of LSMO/Co bilayers without spacers or tunneling barriers, we observe the opposite behavior, that is larger resistance for the increasing-|H| branches and smaller resistance for the decreasing-|H| branches. Also in some LSMO/$Cu_2O$/Co samples with $Cu_2O$ thickness ≤5nm we observe this

inverted hysteresis behavior, which provides a useful warning of electrical shorts. This is an important check which allows us to consider and analyze only short-free samples.

In this respect, it is worthwhile to spend a few words on the possible electrical shorts across the trilayer structures. We think that the not too high resistance of the $Cu_2O$ layer allows us to reasonably ignore this possibility, according to the following argument. An hypothetical spherical particulate which could short the $Cu_2O$ layer should have a diameter equal or larger than the $Cu_2O$ thickness and a resistivity such that its resistance is comparable or smaller than the resistance of the $Cu_2O$ layer, calculated from the resistivity of figure 3 and from the geometrical factors. Hence an upper limit for the particulate resistance is obtained. This value is around $10^{-6}$ Ωm at low temperature, much lower than any reasonable value for binary or simple oxides which could come from the pulsed laser deposition. A large number of more resistive particulate could also short the $Cu_2O$ layer, but no such particulates are observed by AFM imaging. At high temperature, the resistivity of the $Cu_2O$ layer is much smaller, so that the upper limit for the particulate resistance is much smaller and the condition becomes even more safely fulfilled. As an *a posteriori* check, the possibility of electrical shorts is ruled out by comparing measurements on trilayers with different $Cu_2O$ thickness: despite the absolute values of the measured resistances do not scale with the $Cu_2O$ thickness, due to the dominant electrode contribution, a trend of the hysteretic term in the transport data is identified, as described in the following, confirming the pinhole-free behavior of our samples. Only in the case of very thin $Cu_2O$ layers (≤5nm) we find evidence of electrical short, due to the significant $Cu_2O$ roughness and consequent non uniform coverage of the bottom electrode. In this case, we observe a different hysteretic effect (larger resistance for the increasing-|H| branches and smaller resistance for the decreasing-|H| branches), similar to the case of LSMO/Co bilayers, possibly related to the stronger magnetic coupling between the two ferromagnetic electrodes. An independent confirmation that the effective $Cu_2O$ thickness crossed by spin polarized carriers coincides with the macroscopic measured thickness will be obtained by measuring spin diffusion in planar structures, as described in the concluding section.

In figure 8, we show resistance versus cyclically swept magnetic field on LSMO/$Cu_2O$/Co trilayers with different $Cu_2O$ thickness at T=10K, from $t_{Cu2O}=5nm$ (in this case we assume that carriers travel across the $Cu_2O$ layer via tunneling rather than usual transport) to $t_{Cu2O}\approx 125$ nm. Due to the above mentioned series contribution of electrodes to the measured resistance, the magnitude of the hysteretic contribution $\Delta R/R_0=(R^+-R^-)/R_0$ yields a severely underestimated spin polarization, using the Jullière formula [38]. Here, $R^+$ and $R^-$ indicate the resistances of the decreasing-|H| and increasing-|H| branches, respectively, at the field where their difference is the largest, and $R_0$ indicates their average value $(R^++R^-)/2$. However, if we take as a reference the hysteresis $\Delta R/R_0|_{ref}$ of the $t_{Cu2O}\equiv t_{Cu2O}^{ref}=5nm$ structure, where tunneling rather than transport across $Cu_2O$ likely occurs, we can tentatively extract the suppression of the spin polarization across the $Cu_2O$ layer, in the assumption that the series resistance contribution of the electrodes is roughly the same for all the samples. The relative hysteretic contribution $\Delta R/R_0$ is extracted for all the samples at different temperatures (where the curve is not symmetric with respect to the sign of H the average $\Delta R$ is taken) and the results are plotted in the upper panel of figure 9. For thicknesses $t_{Cu2O}\approx 100$ nm or larger, the hysteresis is negligibly small; for thicknesses $t_{Cu2O}\approx 75$ nm or smaller the hysteresis monotonically decreases with increasing thickness and also with increasing temperature, vanishing completely at 300K, where the LSMO layer is too close to its Curie temperature. The spin diffusion length in $Cu_2O$ $d_{spin}$ as a function of temperature can be extracted from the relationship $\left.\frac{\Delta R}{R_0}\middle/\left.\frac{\Delta R}{R_0}\right|_{ref}\right. \approx \exp\left(-\frac{t_{Cu2O}-t_{Cu2O}^{ref}}{d_{spin}}\right)$. An example of such fit on the T=10K data is shown in the lower right panel of figure 9 and the results at different temperatures are plotted in the lower left panel of the same figure. At the lowest temperature $d_{spin}$ is around 40nm. With increasing temperature, $d_{spin}$ decreases weakly. Consistently, also the temperature dependence of the mean free path in $Cu_2O$ is found to be very weak. In actual facts, it is possible that the real temperature dependence of $d_{spin}$ is even weaker than what shown in figure 9; indeed, this temperature dependence is extracted from the $\Delta R/R_0$ data, but it is likely that the decreasing spin polarization of the LSMO electrode contributes to this effect more significantly than the temperature dependence of the $Cu_2O$ spin diffusion length itself.

The above quantitative treatment should be taken with some caution due to the above mentioned limits related to the series resistance of electrodes and the few data points available. However, the monotonic trend of $\Delta R/R_0$ as a function of $t_{Cu2O}$ for the four samples with $t_{Cu2O}\approx5$ nm, $t_{Cu2O}\approx50$ nm, 75 nm and 100 nm and the fact that, apart from the reference $t_{Cu2O}\approx5$ nm sample, all the $Cu_2O$ thickness values are more than one order of magnitude larger than typical tunneling thicknesses indicate unambiguously that a fraction of the carriers that travel across the $Cu_2O$ layer remain spin polarized for distances almost as far as 100 nm. The corresponding spin diffusion length $d_{spin}\approx40$nm, though not close to the record values for high mobility semiconductors [12,14], is almost two orders of magnitude larger than the mean free path of charge carriers. If we assume that the depolarization is due to the Elliot-Yafet mechanism [39], our result indicates that in $Cu_2O$ the spin-orbit scattering Hamiltonian is much smaller than the total scattering potential, as expected for a compound made of light elements. This makes $Cu_2O$ a potentially suitable semiconductor for spin transport applications. We suggest that obtaining $Cu_2O$ samples with larger mean free path could help in improving further the spin diffusion length. This is possible in thinner films and thus in planar devices fabricated with ultrathin $Cu_2O$ films ($t_{Cu2O}$<20nm), possibly deposited on more matched substrates such as MgO. On the other hand, 50-100nm is just the typical size of $Cu_2O$ grains (see figure 2), indicating that grain boundaries may have a crucial role in spin depolarization. Again, the granular structure is strongly improved in thinner films, so that this points as well to the possibility of better performances of $Cu_2O$ planar devices.

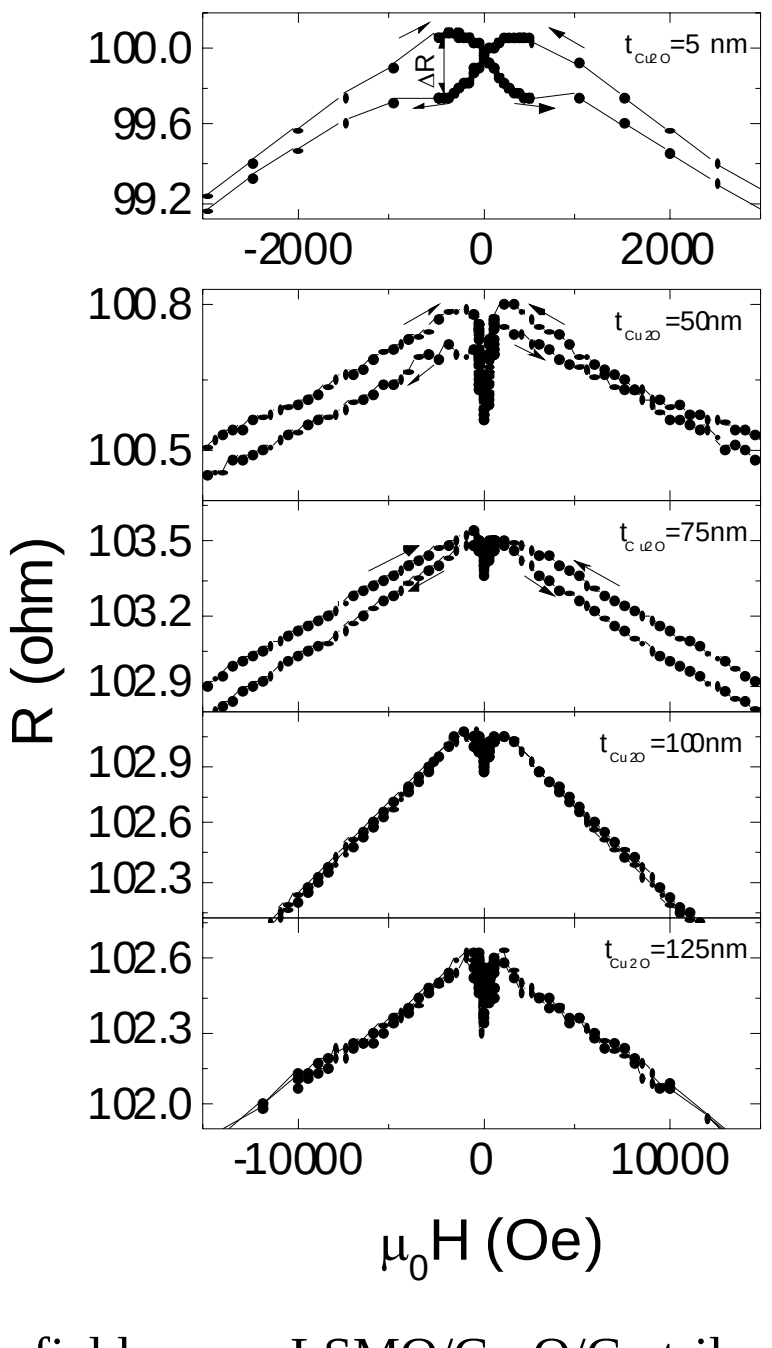

**Figure 8:** Resistance versus magnetic field across LSMO/$Cu_2O$/Co trilayers with different $Cu_2O$ thicknesses, measured at T=10K. The arrows indicate the direction of the magnetic field sweep. Our definition of the hysteretic contribution $\Delta R$ is indicated in the top panel curve. The horizontal axis of the top panel is zoomed to better emphasize the hysteresis; indeed, being the $Cu_2O$ spacer only 5nm thick, the residual magnetic coupling between the top and bottom electrodes causes the hysteresis to be visible only in a smaller range of magnetic fields.

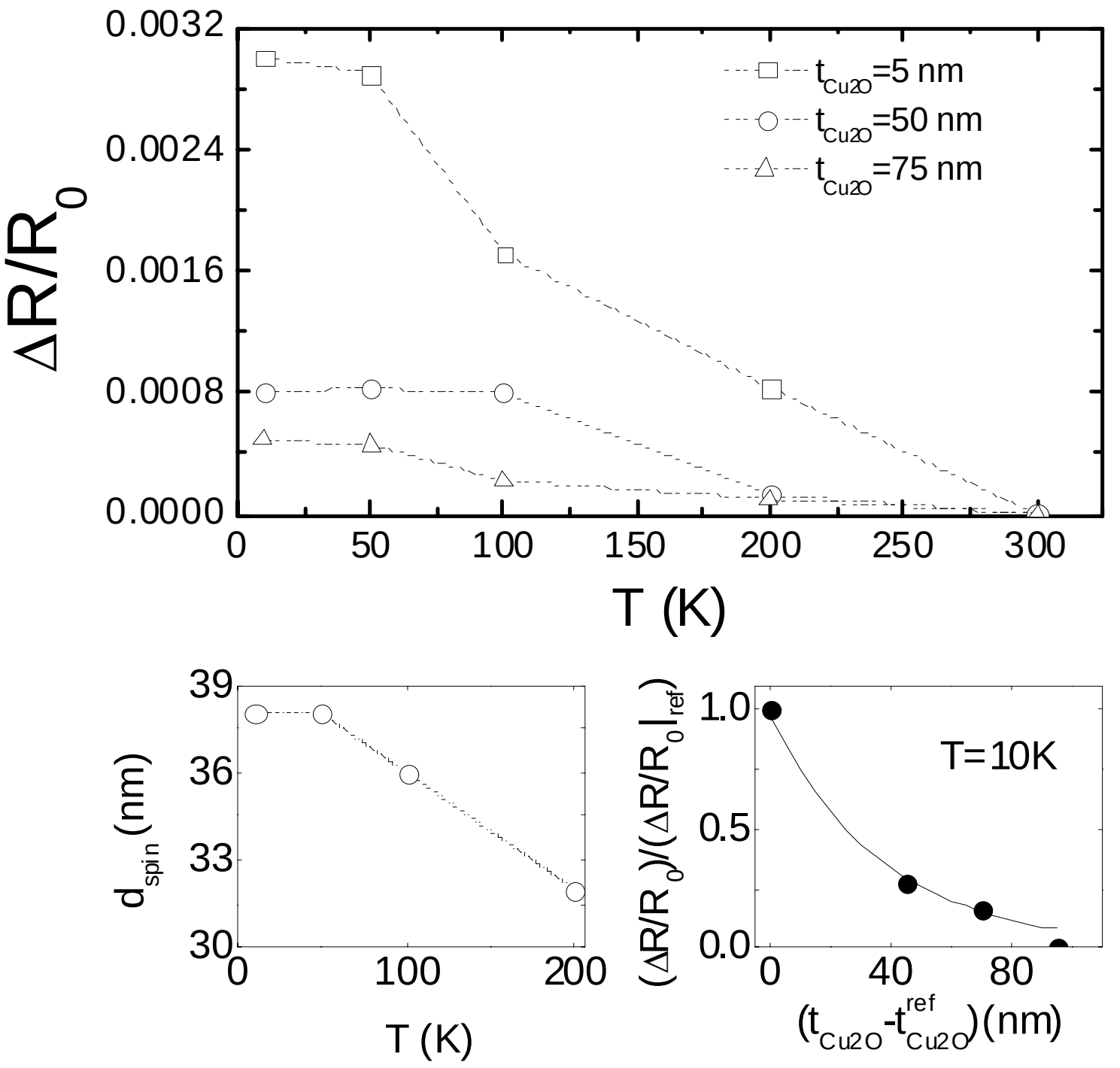

**Figure 9:** Upper panel: hysteretic contribution $\Delta R/R_0=(R^+-R^-)/R_0$ to magnetoresistivity of LSMO/$Cu_2O$/Co heterostructures with different $Cu_2O$ thicknesses, where $R^+$ and $R^-$ the resistances of the decreasing-|H| and increasing-|H| branches, respectively, and $R_0$ their average value. Lower left panel: extracted spin diffusion length in $Cu_2O$ inside LSMO/$Cu_2O$/Co heterostructures as a function of temperature. Lower right panel: exponential fit of $\left.\frac{\Delta R}{R_0}\middle/\frac{\Delta R}{R_0}\right|_{ref}$ data at T=10K as a function of $(t_{Cu2O}-t_{Cu2O}^{ref})$.

We finally address the effect of a $SrTiO_3$ tunneling barrier between LSMO and $Cu_2O$ layers, which could help inferring information about band alignment at the LSMO/$Cu_2O$ interface. Indeed, spin valves can be classified into two categories, depending on the interface resistance [15,16]: on one side are the tunnel junctions [40], whose interface resistance is much larger than the characteristic spin resistance (i.e. the interface resistance related to the voltage drop caused by spin accumulation) of the nonmagnetic material, on the other side are the transparent junctions [8], where the opposite situation occurs. In the former case there is negligible interaction between the magnetic and normal materials and the spin accumulation in the nonmagnetic material decays exponentially with the distance from the interface, while in the latter case depolarization is influenced by relaxation processes occurring not only in the nonmagnetic material but also in the ferromagnetic electrode. As a consequence, larger spin valve effects are obtained in tunnel junctions, which in turns suffer of limited spin current density and bias dependent depolarization. Transparent junctions, although liable to larger depolarization at the interface, can transfer efficiently spin currents, thanks to the largest sustainable current density [41].

We are unable to prepare a LSMO/$SrTiO_3$/$Cu_2O$/$SrTiO_3$/Co heterostructure with two crystalline $SrTiO_3$ barriers, because the upper $SrTiO_3$ layer cannot be obtained in the crystalline form, as explained in the experimental section. On the other hand, an amorphous $SrTiO_3$ is detrimental for spin polarization across the interface. Hence, only LSMO/$SrTiO_3$/$Cu_2O$/Co heterostructures are analyzed in the following. The current-voltage characteristics (not shown) of all the heterostructures with and without the $SrTiO_3$ tunneling barriers are ohmic. We do not have a complete series of samples whose only difference is the thickness of the $SrTiO_3$ layer $t_{SrTiO3}$, thereby in figure 10 we show separately the effects of a 5nm $SrTiO_3$ barrier in a structure with $t_{Cu2O}=50nm$ (upper panel) and of 10nm and 15nm $SrTiO_3$ barriers in structures with $t_{Cu2O}=5nm$ (lower panel). In all cases, the curves are compared with the respective reference curves measured on LSMO/$Cu_2O$/Co heterostructures having the same $Cu_2O$ thickness ($t_{Cu2O}$=50nm in the upper panel and $t_{Cu2O}$=5nm in the lower panel). We display data at a specific temperature, T=100K, as an example. The curves are normalized to the value $R_0$ defined as above, for better visualization and comparison. It can be seen that the 5nm $SrTiO_3$ barrier has no detectable effect on the hysteretic behavior ($\Delta R/R_0$~0.0009 for $t_{Cu2O}$=50nm at T=100K for both LSMO/$SrTiO_3$/$Cu_2O$/Co and LSMO/$Cu_2O$/Co heterostructures), while with increasing barrier thickness $\Delta R/R_0$ decreases with respect to the reference $\Delta R/R_0$ of the $SrTiO_3$-free sample. All $\Delta R/R_0$ data extracted from the plots of figure 10 and normalized to the corresponding $SrTiO_3$-free sample values are summarized in the inset of the same figure. The result indicates that the $SrTiO_3$ barrier is not effective in improving spin injection in these systems. It is possible that LSMO forms naturally a barrier at the interface [13,42], commonly called "dead layer" [43], which acts as a barrier itself. However, a more plausible explanation for the negligible effect of the additional $SrTiO_3$ barrier is that the LSMO/$Cu_2O$ interface does not suffer of the problem of resistivity mismatch, as both materials have fairly high resistivities as compared to the typical metallic values. The similar behaviour of LSMO/$SrTiO_3$/$Cu_2O$/Co and LSMO/$Cu_2O$/Co heterostructures also indicates that the most critical interfaces for spin depolarization are the perovskite/$Cu_2O$ ones or else the "dead layer" at the LSMO interface, which are present in both types of heterostructures, so that the insertion of the additional LSMO/$SrTiO_3$ and $SrTiO_3$/$Cu_2O$ interfaces has a minor effect.

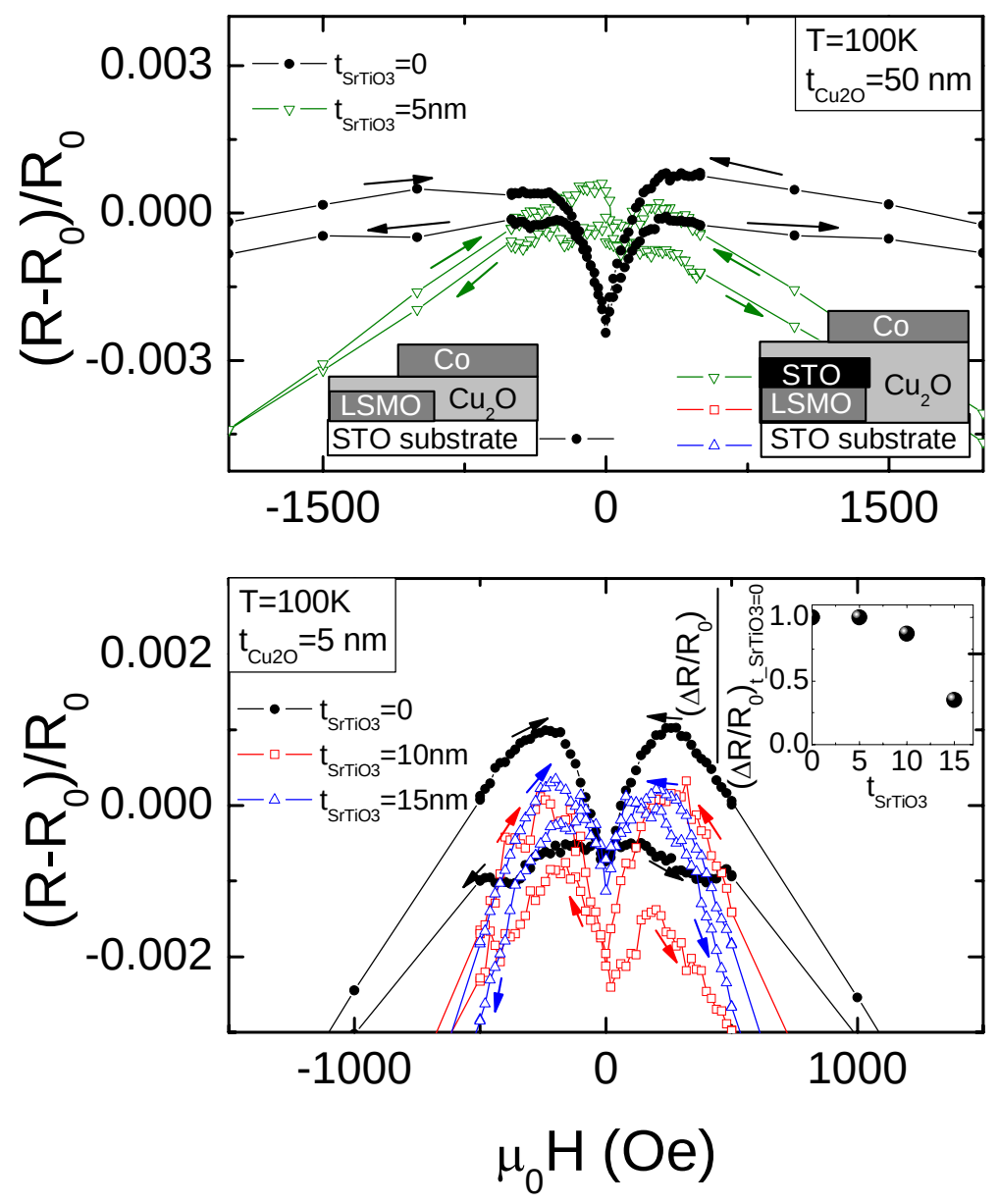

**Figure 10:** (color online) Upper panel: resistance versus magnetic field across LSMO/$Cu_2O$/Co and LSMO/$SrTiO_3$(5nm)/$Cu_2O$/Co trilayers with $t_{Cu2O}=50nm$, measured at T=100K and normalized to the average resistance value $R_0=(R^++R^-)/2$ (see text for definition of $R^+$ and $R^-$). In both cases a comparable value $\Delta R/R_0\approx 9\times 10^{-4}$ is observed. Lower panel: same as above but for trilayers with with $t_{Cu2O}=5nm$ and different $t_{SrTiO3}=10nm$ and $15nm$. In both panels, the arrows indicate the direction of the magnetic field sweep. In the inset of the lower panel, the $\Delta R/R_0$ data extracted from the curves in the main panels and normalized to the corresponding $SrTiO_3$-free sample values are plotted. The device layer sequences LSMO/$Cu_2O$/Co and LSMO/$SrTiO_3$/$Cu_2O$/Co are sketched in the insets of the upper panel, close to the corresponding symbols of the data plots in both upper and lower panels.

## 4. Conclusions and developments

We probe spin transport in $Cu_2O$ by measuring spin valve effect in LSMO/$Cu_2O$/Co and LSMO/$Cu_2O$/LSMO trilayers. We extract an estimation of the spin diffusion length in $Cu_2O$, which is around 40 nm at low temperature. This value is two orders of magnitude larger than the charge mean free path and also much larger than all the characteristic length scales of correlated oxides, whose screening length for example is few nm at most. Oxide electronics, based on crystalline integration of functional materials could overcome some limits of Silicon based electronics just thanks to the smaller characteristic lengths into play, which could allow to shrink the device size. Indeed, we find that some fraction of spin polarization survives up to distances of almost 100nm across $Cu_2O$, which is a length scale compatible with currently available nanopatterning technologies. Hence, it appears that $Cu_2O$ is a potentially suitable material for spin transport in sub-micrometric oxide electronic devices.

Both mean free path and spin diffusion length in $Cu_2O$ films may be improved by optimizing its structural quality and physical properties. For example the use of a planar geometry based on $Cu_2O$/Co bilayers is more versatile in terms of $Cu_2O$ thickness, allows to grow $Cu_2O$ on MgO substrates, which have more matched lattice parameters with $Cu_2O$ than $SrTiO_3$ [20], and is intrinsically less liable to electrical shorts than stacked layer structures. Furthermore, $Cu_2O$/Co structures allow spin valve measurements up to room temperature and even above, thanks to the high Curie temperature of Co. However, once spin transport in $Cu_2O$ is fully explored and its limits identified, a full-crystalline-oxide spin injection device will be the ultimate target of this investigation.